\documentclass[twocolumn,english,aps,prb,showpacs]{revtex4}

\usepackage{graphicx}
\usepackage{epsfig}
\begin{document}

\title{Superconductivity in $\beta$-pyrochlore superconductor KOs$_{2}$O$_{6}$: treatment within strong-coupling
Eliashberg theory}

\author{Jun Chang$^{1 \footnote{Also at Institute of Theoretical Physics, Chinese Academy of
Sciences,Beijing 100080, China}}$, Ilya Eremin$^{1,2}$, and Peter
Thalmeier$^{3}$}

\affiliation{$^{1}$ Max-Planck-Institut f\"{u}r Physik komplexer Systeme, D-01187 Dresden, Germany \\
 $^{2}$ Institute f\"{u}r Mathematische und Theoretische Physik, TU-Braunschweig, D-38106 Braunschweig, Germany \\
 $^{3}$ Max-Planck-Institut f\"{u}r Chemische Physik fester Stoffe, D-01187 Dresden, Germany}

\date{\today{}}

\pacs{74.20.-z, 74.25.Kc, 74.20.Fg}

\begin{abstract}
We study the influence of the rattling phonons on superconductivity
in $\beta$-pyrochlore KOs$_{2}$O$_{6}$ compound based on the
strong-coupling Eliashberg approach. In particular, analyzing the
specific heat data we find that the rattling phonon frequency
changes discontinuously at the critical temperature of the first
order phase transition.  Solving the strong-coupling Eliashberg
equations with temperature dependent $\alpha^{2}F(\omega)$, we
investigate the consequence of this first order phase transition for
the anomalous temperature dependence of the superconducting gap. We
discuss our results in context of the recent experimental data.
\end{abstract}
\maketitle

The superconductivity in the family of the $\beta$-pyrochlore oxides
KOs$_2$O$_6$, RbOs$_2$O$_6$, and CsOs$_2$O$_6$ exhibits many exotic
properties\cite{Yonezawa1,Yonezawa2,Bruhwiler1,Magishi,Khasanov,Arai}.
Among them, KOs$_2$O$_6$ with superconducting transition temperature
T$_c=9.6$K shows the most anomalous behavior. For example, the
electrical resistivity demonstrates strong concave $T$ dependence
down to T$_c$\cite{Yonezawa1}, in contrast to the normal $T^2$
behavior in Rb and Cs
compounds\cite{Yonezawa2,Bruhwiler1,Magishi,Khasanov,Arai}. The
specific heat measurements have found an existence of low frequency
Einstein modes and that the $T$-linear coefficient of the specific
heat $\gamma = 70 mJ/K^2 \cdot mol$, (see Refs.
\onlinecite{Bruhwiler2,Hiroi}), is strongly enhanced over the value
obtained from band structure calculations\cite{Kunes,Saniz}. The
band structure calculations have indicated that these anomalies may
be due to highly anharmonic low frequency rattling motion of the
alkali-ions inside an oversized cage formed by the Os and O
ions\cite{Kunes}. Furthermore, this is consistent with the $x$-ray
observation of anomalously large atomic displacement for the $K$
ions\cite{Yamaura}, and the low frequency phonon structures seen in
photoemission spectra\cite{Shimojima}. Moreover, recent NMR
data\cite{Yoshida} have indicated that the relaxation at the K sites
is entirely caused by fluctuations of the electric field gradient
ascribed to the highly anharmonic low frequency dispersionless
oscillation (rattling) of K ions in the cage. Recently, Dahm and
Ueda have developed the phenomenological model to describe the
influence of the anharmonic phonons on the NMR relaxation and the
electrical resistivity\cite{Dahm}.

Various experiments have been carried out in KOs$_2$O$_6$ to
elucidate the mechanism of the superconductivity. Most importantly,
thermal conductivity\cite{kasahara} and laser photoemission
spectroscopy(PES) \cite{Shimojima} measurements have revealed almost
isotropic $s$-wave superconducting gap. This means that the
Cooper-pairing can be ascribed to phonons. Furthermore, an absence
of the coherence peak in the K-NMR relaxation rate can be explained
due to strongly overdamped phonons\cite{Yoshida}. At the same time,
a large value of the electron-phonon coupling constant $\lambda_{ep}
\approx 2.38$, the presence of the low-energy rattling phonon modes
and their influence on the electronic properties, and finally, the
first-order structural phase transition at T$_p <$ T$_c$
\cite{Hiroi} makes an investigation of superconductivity in
KOs$_2$O$_6$ very interesting, in particular, from a point of view
of studying further aspects of the strong coupling Eliashberg theory
in presence of coupling between conduction electrons and rattling
phonons. Previously, an attempt to consider the effect of the
temperature-independent rattling phonons has been considered in
Ref.\onlinecite{battlog} in application to CsOs$_2$O$_6$.

In this Brief Report, we analyze the superconductivity in
KOs$_2$O$_6$ using the standard Eliashberg formalism supplemented by
a quasiharmonic treatment of the rattling mode and its
renormalization by the coupling with conduction electrons. We find
that the phonon spectrum can be modeled by the two Lorentzians
peaked at the energies of the two Einstein modes, $\omega_{E1}$ and
$\omega_{E2}$ representing the lowest energy rattling phonons. An
additional contribution arises from the Debye frequency at
$\omega_D$. To explain the superconducting transition temperature in
KOs$_2$O$_6$ we have employed the mean-field analysis of the
temperature dependence of the lowest Einstein mode proposed
previously\cite{Dahm}. We show that its energy should decrease with
temperatures down to T$_p < $ T$_c$ and then jump to a higher
frequency indicating the first order structural phase transition.
Solving the non-linear Eliashberg equations below T$_c$ with
temperature dependent electron-phonon coupling function, $\alpha^2
F(\omega)$ we compare our results with various experiments.

On the real frequency axis the finite temperature Eliashberg
equations for the superconducting gap $\Delta (\omega, T)$ and the
renormalization function $Z(\omega, T)$  are given by
\cite{holcomb}:
\begin{eqnarray}
\Delta(\omega,T) &=& \frac{1}{Z(\omega,T)} \int_{0}^{\infty}
d\omega' \mbox{Re} \left\{
\frac{\Delta(\omega',T)}{\sqrt{\omega'^{2}-\Delta^{2}(\omega',T)}}\right\}\nonumber\\
&\times&\left[K_{+}(\omega,\omega',T)-\mu^{*} \tanh
\left(\frac{\beta\omega'}{2} \right) \right] \quad,
\end{eqnarray}
\begin{eqnarray}
\omega(1-Z(\omega,T)) &=& \int_{0}^{\infty} d\omega' \mbox{Re}
\left\{
\frac{\omega'}{\sqrt{\omega'^{2}-\Delta^{2}(\omega',T)}} \right\}\nonumber\\
&\times&K_{-} (\omega,\omega',T) \quad,
\end{eqnarray}
where
\begin{eqnarray}
\lefteqn{K_{\pm}(\omega,\omega',T) = \int_{0}^{\infty} d\Omega \,
\alpha^2 F(\Omega)
\left[ \frac{f(-\omega')+n(\Omega)}{\omega' + \omega + \Omega} \right.} && \nonumber\\
&& \left. \pm \frac{f(-\omega')+n(\Omega)}{\omega' - \omega +
\Omega} \mp \frac{f(\omega')+n(\Omega)}{-\omega' + \omega + \Omega}
- \frac{f(\omega')+n(\Omega)}{-\omega' - \omega + \Omega} \right],
\nonumber\\
\end{eqnarray}
and
\begin{equation}
\alpha^2
F(\Omega)=\alpha_{E1}^{2}F_{E1}(\Omega)+\alpha_{E2}^{2}F_{E2}(\Omega)+\alpha_{P}^{2}F_{P}(\Omega)
\quad,
\end{equation}
is the generalized electron-phonon coupling function averaged over
the Fermi surface. We assume that there are two contributions to the
Eliashberg function. The second is arising from the usual dispersive
acoustic phonons limited by the Debye energy $T_{D}=325$ K. The
first contribution is due to local low energy rattling phonons whose
energies we denote by $\omega_{E1}$ and $\omega_{E2}$. The latter
requires a special consideration.

It has been shown recently that the low-energy phonons ascribed to
the heavy rattling of the K ions confined in an oversized cage made
of OsO$_6$ octahedra are responsible for the unusual scattering
processes in KOs$_2$O$_6$\cite{Hiroi}. In particular, it has been
assumed that contrary to the RbOs$_2$O$_6$ and CsOs$_2$O$_6$ cases
where a single Einstein mode due to rattling motion is enough, there
exits two modes at $\hbar \omega_{E1}=22$K, and $\hbar
\omega_{E2}=61$K in KOs$_2$O$_6$ corresponding to the first two
excited energy levels of the corresponding anharmonic potential.
Simultaneously, the mean-field description of the local alkali-ion
anharmonic motion has been developed in Ref. \cite{Dahm}. Starting
from the standard anharmonic Hamiltonian
\begin{equation}
H=\frac{p^{2}}{2M}+\frac{1}{2}ax^{2}+\frac{1}{4}bx^{4} \quad,
\label{eq:ph}
\end{equation}
where $x$, $p$, and $M$ are the spatial coordinate, momentum, and
mass of the alkali ion, respectively and $a,b$ are constants with
$b>0$ one finds that according to {\it ab initio} calculations
\cite{Kunes} for KOs$_{2}$O$_{6}$, the quadratic term becomes
negative, {\it i.e.} $a<0$, that results in a double well potential.
We note that in reality the double wells are not in the
plane\cite{Kunes}. Treating the Hamiltonian (\ref{eq:ph}) in the
self-consistent quasiharmonic approximation Dahm and Ueda have found
that the oscillation of Potassium ions can be described by an
effective harmonic oscillation with effective low-energy frequency
that now depends on temperature\cite{Dahm}. In particular, it
decreases monotonically upon decreasing temperature\cite{remark1}.
At low enough temperatures ($T\ll \omega_0$), this frequency becomes
nearly temperature independent. Obviously this model requires
certain modification when applied to KOs$_2$O$_6$. In particular,
although this effective model works well at high temperatures, it
fails to explain the occurrence of the first order phase transitions
at temperature $T_p < T_c$ and the occurrence of the second Einstein
frequency at higher energy.

Therefore, modeling the contribution of the rattling K ions to the
Eliashberg function we adopt two Einstein like modes centered at
$\omega_{E1}$ and $\omega_{E2}$. Furthermore, to take into account
the upper lying energy levels of the shallow potential we use the
self-consistent quasiharmonic approximation for the lowest mode,
$\omega_{E1}$ that results in its temperature dependence. Most
importantly, before entering the temperature independent regime the
effective rattling phonon frequency shows a discontinuous jump
towards higher frequency that will result in the specific heat
anomaly at $T_{p}$. The increase of the $\omega_{E1}$ at T$_p$ is in
accordance with recent experiments showing no structural changes
below the phase transition\cite{Hiroi,Yoshida,Hasegawa}. Indeed,
below T$_p$ the K ions seem to be stabilized in their equilibrium
positions which corresponds to the increase of $\omega_{E1}$.
\begin{figure}
\includegraphics[width=0.8\columnwidth]{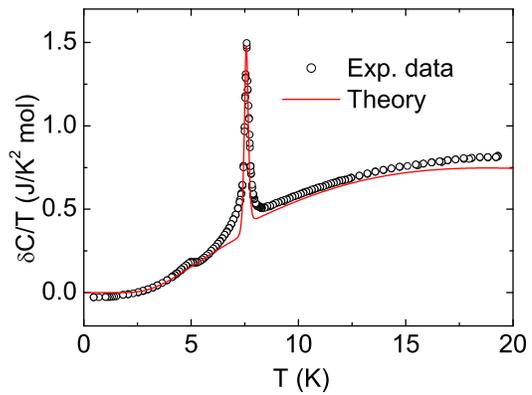}
\caption{(color online) Theoretical fit to the experimental specific
heat data from Ref. \onlinecite{Hiroi} using $\hbar \omega_{E2} =
59$K and $\hbar \omega_{E1}(T=T_c) = 27.4$K. Assuming the jump of
the characteristic rattling phonon frequency at T$_p$ from $\hbar
\omega_{E1}$=27.4K to $\hbar \omega_{E1}$=30K  we find the
correspondent increase in the specific heat curve in qualitative
agreement with experiment data. } \label{fig-heat}
\end{figure}

\begin{figure}
\includegraphics[width=0.8\columnwidth]{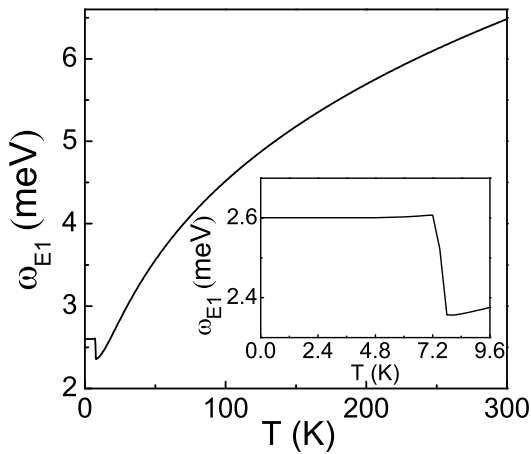}
\caption{Calculated temperature dependence of the lower frequency,
$\omega_{E1}$ of the rattling phonon  as obtained using the
quasi-harmonic approximation\cite{Dahm}. The parameter of the model
$b\hbar / M^2 \omega^3_{0}(T=300K)=0.25$ have been fixed by assuming
the room temperature position of the rattling phonon frequency
around $6$ meV as found by neutron scattering
experiments\cite{hiroi_neutron}. At T$_p$ the effective frequency
demonstrates the discontinuous jump to lower energy (as shown in the
inset) which position is determined by the absolute value of the
specific heat jump shown in Fig.\ref{fig-heat}. } \label{fig-omegaT}
\end{figure}

To find the exact shift of the lowest frequency at T$_p$, and also
the approximate positions of the rattling mode frequencies we made a
simulation of the specific heat data around T$_p$ following the
model proposed by Ref. \onlinecite{Hiroi} with temperature dependent
$\omega_{E1}$. The contribution from the two frequencies
representing the rattling modes is given by\cite{Hiroi}
\begin{equation}
C =  aC_{E1} +(1-a)C_{E2} \label{eq:heat}
\end{equation}
where $a=0.24$ and $\omega_{E_i}$  with $i=1,2$ are the
corresponding Einstein phonons. As mentioned above, for the sake of
simplicity we assume that the higher mode is temperature independent
and the lower one follows the mean-field temperature dependence that
originates from integrating out the upper-lying energy levels with
temperature. Generally, for the contribution of temperature
dependent Einstein frequency one finds
\begin{equation}
C_{E}=3R \left(\frac{\hbar \omega_E}{k_B T}\right)^2
\frac{\exp\left(\frac{\hbar \omega_E}{ k_B
T}\right)}{\left(\exp\left(\frac{\hbar \omega_E}{ k_B T}\right)
-1\right)^2} \left[ 1 - \frac{\partial \ln \omega_E}{ \partial \ln
T} \right]
\end{equation}
where $R$ is the gas constant. The last term is absent for the upper
temperature-independent frequency, $\omega_{E2}$. The results of the
fit are shown in Fig.~\ref{fig-heat}. The change of the specific
heat at $T_p$ is reproduced by assuming the jump of the lower
Einstein frequency from $\hbar \omega_{E1}$=27.4K to $\hbar
\omega_{E1}$=30K at this temperature with $\omega_{E2}$ being
constant. Above T$_p$ the lower frequency mode is temperature
dependent and the overall behavior of $\omega_{E1}(T)$ is shown in
Fig.\ref{fig-omegaT}. Note that the strong temperature dependence of
the lowest rattling phonon mode frequency has been recently observed
by the Raman scattering\cite{Hasegawa,schoenes}.

In addition to the anharmonic temperature dependence of the rattling
phonon mode which modifies the usual approximation for the
Eliashberg function, the conduction electrons couple to the rattling
phonons, further renormalizing its energy position and introducing
an extra damping. In particular, the spectral function of the
low-energy rattling phonon frequency is given by
\begin{equation}
\alpha^2 F_{E1}(\omega)=-\frac{\alpha^2(T)}{\pi}\mbox{Im}D(\omega)
=\frac{\alpha^2(T)}{\pi}\frac{4\omega_{E1}(T) \Gamma_0
\omega}{\left(\omega^{2}-\omega_{r}^{2}\right)^2+4\Gamma_0^2
\omega^2} \quad,
\end{equation}
where $\Gamma_{0}$ is a anharmonic phonon damping rate. The real
part of the rattling phonon self-energy leads to a renormalized
phonon frequency,
\begin{equation}
\omega_{r}^{2}(T)=\omega_{E1}^{2}(T)+2\omega_{E1}(T)\mbox{Re}\Pi(\omega)
\quad.
\end{equation}
Following previous estimation\cite{Dahm} we have used
-Re$\Pi(\omega)=1$ meV and $\Gamma_0=0.25$ meV.  It is important to
remember that the electron-phonon coupling parameter $\alpha^2(T)
\sim c/\omega_{E1}(T)$ with $c=1.23$meV$^2$ is also temperature
dependent.
\begin{figure}
\includegraphics[ width=0.8\columnwidth]{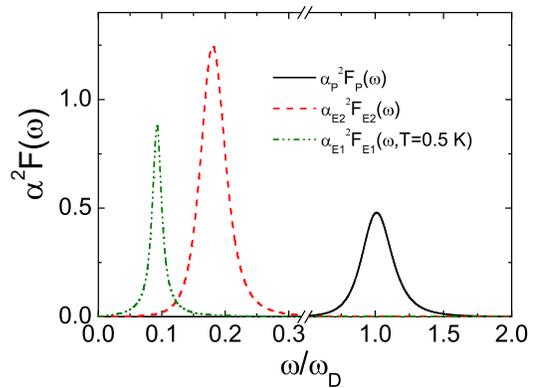}
\caption{(color online) Calculated Eliashberg function, Eq.(4), for
KOs$_{2}$O$_{6}$ for T=0.4K. The solid curve denotes the usual
phononic contribution centered around $\Omega_D$ and the (red)
dashed curve refer to the upper rattling phonons frequency centered
around $\omega_{E2}$. Here, we employ $\alpha^2 F_i(\omega) =
\frac{c_{i} \omega_{i}
\Gamma_i^3}{\pi\left(\left(\omega-\omega_{i}\right)^2+\Gamma_{i}^2\right)^2}$
where $i=E2,p$. We further use $\Gamma_i=\omega_{i}/5$, $c_{E2} =
0.767$, and $c_{p} = 0.3$. The (green) dashed-dotted curve refers to
the lower energy rattling mode contribution to $\alpha^2 F (\omega)$
with parameters as described in the text.} \label{fig-spectra}
\end{figure}

\begin{figure}[t]
\includegraphics[ width=0.8\columnwidth]{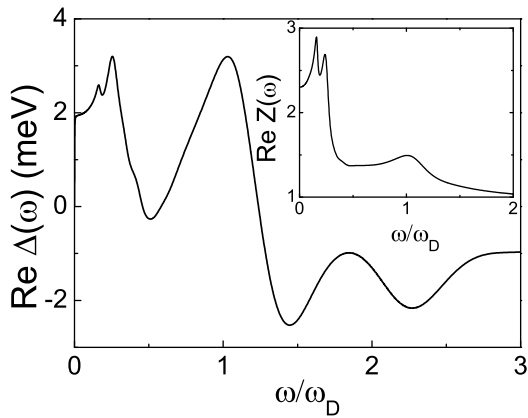}
\caption{Calculated frequency dependence of the real part of the
superconducting gap function $\Delta(\omega)$ for KOs$_{2}$O$_{6}$
compounds. The inset shows the renormalization function, $Z(\omega)$
at $T=0.5$K. Note that we set the cutoff frequency to $8\omega_{D}$
and introduce a finite damping $\Gamma_0=0.25$ meV.}
\label{fig-DeltaZ}
\end{figure}

In Fig.\ref{fig-spectra} we show the Eliashberg function $\alpha_i^2
F_i(\Omega)$ including all three contributions for T$=0.5$K {\it
i.e.} well below T$_p$. In contrast to the usual Eliashberg theory
the $\alpha^2 F(\omega)$ spectrum and the coupling constant
$\lambda=2\int_{-\infty}^{+\infty}d\Omega \alpha^2 F(\Omega)/\Omega$
are now temperature dependent. Setting $\mu^{*}=0.091$ and solving
the Eqs. (1)-(3) in the linearized limit we find the superconducting
transition temperature T$_c$=9.6K. In Fig. \ref{fig-DeltaZ} we show
the results of the solution of the Eliashberg equations for
$T=0.5$K. One finds the typical behavior of a strong-coupling
superconductor. In particular, Re$\Delta(\omega)$ shows three peaks
at the energy of the rattling phonons and at $\omega_D$. For larger
energies it becomes negative reflecting the effective repulsion for
$\omega>\omega_D$. Due to presence of the low-energy rattling
phonons one obtains a strong renormalization of the quasiparticle
mass and we also find $2\Delta_{0}/k_{B}T_{c}\approx$ 5.0. The
latter is in good agreement with experimentally observed
value\cite{Shimojima}.

\begin{figure}[t]
\includegraphics[width=0.8\columnwidth]{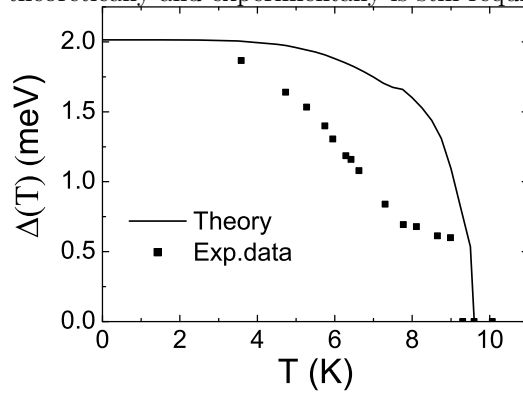}
\caption{Calculated temperature dependence of the superconducting
gap $\Delta_0(T)$ as determined from Re$\Delta(T, \omega)=\omega$
for KOs$_{2}$O$_{6}$. The black squares show the experimental data
from Ref. \onlinecite{Shimojima} as obtained from the fit of the
experimental curve by the Dynes formula.} \label{fig-DeltaT}
\end{figure}

One of the important consequences of the temperature dependent
phonon spectrum is slightly anomalous behavior of the
superconducting gap as a function of temperature. In
Fig.\ref{fig-DeltaT} we show the temperature dependence of the
superconducting gap. One finds a kink at T=T$_p$ due a discontinuous
change of the phonon frequency and the corresponding coupling
constant. In particular, due to the slight increase of $\omega_{E1}$
below T$_p$ the electron-phonon coupling constant decreases which
yields lowering of the superconducting gap with respect to its value
without the structural phase transition. At the same time, we find
overall relatively small modification of the superconducting
properties due to structural phase transition. This is consistent
with experimental observation\cite{Hiroi} which show weak connection
between the structural transition and superconductivity in
KOs$_2$O$_6$. The most important result is that due to temperature
dependent low-energy mode the superconductivity is enhanced in
KOs$_2$O$_6$ compared to its Rb and Cs counterparts. We also find
that the electron-phonon coupling strength changes from
$\lambda(T>T_p) =1.7$ towards $\lambda(T<T_p) =1.6$ and is somewhat
lower than estimated previously\cite{Hiroi}. This is because a
simple application of the McMillan-Dynes formula for estimating
$\lambda$ is probably questionable in KOs$_2$O$_6$ due to
temperature dependence rattling mode which does not exist in simple
Eliashberg theory. Our obtained results although showing an anomaly
at T$_p$ cannot explain fully the photoemission data of
Ref.\cite{Shimojima} as clearly visible from the
Fig.\ref{fig-DeltaT}.

Our calculation shows that the superconducting properties of
KOs$_2$O$_6$ should be sensitive to the external pressure. Due to
proximity to the first-order phase transition the application of the
pressure may result in a modification of the electron-phonon
coupling strength. Such an unusual behavior has been indeed recently
found experimentally \cite{miyoshi} although the detailed
understanding both theoretically and experimentally is still
required.

In summary, we have investigated the superconducting properties of
$\beta$-pyrochlore KOs$_{2}$O$_{6}$ compound based on the
strong-coupling Eliashberg approach. Analyzing the specific heat
data we find that rattling phonon frequency changes discontinuously
at the critical temperature of the first order phase transition.
Solving the strong-coupling Eliashberg equations with temperature
dependent $\alpha^{2}F(\omega)$, we discuss the consequence of this
first order phase transition for the anomalous temperature
dependence of the superconducting gap. In particular, we have found
that the superconducting gap as a function temperature is anomalous
reflecting the temperature dependence of the Eliashberg function.

We acknowledge the fruitful discussions with P. Fulde, T. Dahm, Z.
Hiroi, M. Koza, M. Korshunov, and J. Schoenes.


\begin{thebibliography}{99}
\bibitem{remark1} It is important to remember that the effective
frequency, $\omega_0$, is a thermodynamical average frequency and
should not be mixed with the descrete, temperature idependent energy
levels of the original Hamiltonian\cite{dahm_sns}.
\end{thebibliography}
\end{document}